\documentclass[12pt]{article}

\newcommand\lsim{\mathrel{\rlap{\lower4pt\hbox{\hskip1pt$\sim$}}
    \raise1pt\hbox{$<$}}}
\newcommand\gsim{\mathrel{\rlap{\lower4pt\hbox{\hskip1pt$\sim$}}
    \raise1pt\hbox{$>$}}}

\usepackage{graphicx}
\usepackage{amsfonts}

\begin{document}

\sf
\centerline{\Huge Minimal SO(10) splits supersymmetry}

\vspace{7mm}

\centerline{\large
Borut Bajc
, Ilja Dor\v sner
and Miha Nemev\v sek
}
\vspace{1mm}
\centerline{
{\it\small J.\ Stefan Institute, 1000 Ljubljana, Slovenia}}

\vspace{5mm}
   
\centerline{\large\sc Abstract}
\begin{quote}
\small

A good fit of the fermion masses and mixings has been found in 
the minimal renormalizable supersymmetric SO(10). This solution needs a 
strongly split supersymmetry breaking scenario with gauginos and higgsinos 
around 100\,TeV, sfermions close to $10^{14}$ GeV and a low GUT scale 
of around $6\times 10^{15}$\,GeV. We predict fast proton decays through SO(10) type of $d=6$ 
operators and the leptonic mixing angle $\sin{\theta_{13}}\approx 0.1$.


\end{quote}
\rm

\section{Introduction}

Assuming renormalizability and no other symmetries, the supersymmetric 
SO(10) with three generations of matter $16_F$ and the Higgs representations 
of $210_H$, $126_H$, $\overline{126}_H$ and $10_H$ \cite{Clark:1982ai,Aulakh:1982sw}
turns out to have only $26$ free parameters (on top of the usual soft supersymmetry 
breaking terms) \cite{Aulakh:2003kg} and thus can be considered the minimal 
prototype grand unified theory (GUT) model.

Several efforts have been employed in trying to fit the fermion masses in 
this minimal SO(10) \cite{Babu:1992ia, Brahmachari:1997cq,Matsuda:2000zp,
Matsuda:2001bg,Fukuyama:2002ch,
Goh:2003sy,Goh:2003hf,Bertolini:2005qb,Babu:2005ia}. 
It has been finally found out that due to constraints 
from the Yukawa and Higgs sectors either the neutrino mass scale came out too 
small or the gauge coupling constants entered the non-perturbative regime 
\cite{Aulakh:2005bd,Bajc:2005qe,Aulakh:2005mw,Bertolini:2006pe}. 

Two main objections can be raised to most of the above works. 

First, the gauge and Yukawa couplings at the GUT scale were obtained from the 
assumption of a desert from the low energy supersymmetry (SUSY) scale to the 
GUT scale. This is not always true, since, depending on the various 
Higgs couplings of the model, there could be states lying one or two 
orders of magnitude below the GUT scale. In a scenario with small 
SO(10) representations \cite{Babu:1998wi,Blazek:1999ue,
Dermisek:1999vy,Albright:2000sz} this would always be just 
a small perturbative correction, but in the minimal SO(10) the large number 
of light remnants of the $210_H$, $126_H$ or $\overline{126}_H$ representations 
could make the difference. Although in some cases it has been found that the relevant 
threshold corrections are negligible \cite{Aulakh:2004hm}, this has not been studied
systematically.

Second, the unknown values of the soft terms in the minimal supersymmetric standard model 
(MSSM) make any prediction of the 
fermion masses and mixings impossible. Due to (in principle) large 
finite threshold corrections at the low supersymmetry breaking scale 
\cite{Hall:1985dx,Hall:1993gn} (the possible impact of these effects 
for renormalizable SO(10) has been recently studied in \cite{Aulakh:2008sn}) 
it is in general difficult to find any fitting reliable. 
Even if a good fit were found, unknown soft terms would change it. 
So to have a really predictive model one would need to know how 
supersymmetry is broken and mediated. Of course there are many such models on the 
market, but their construction is usually completely orthogonal to the 
GUT one is considering (an opposite attempt to construct such a supersymmetry 
breaking model in a GUT has been recently given in \cite{Bajc:2008vk}). Also, 
due to the same reason, $d=5$ proton decay is not predicted by the 
theory, although it will typically be dangerously large. In short, 
unless the finite threshold corrections are known to be negligible for 
some reason, there is little hope to have a grand unified 
theory of fermion masses and proton decay.

Here we will show that a consistent treatment of the first issue automatically 
pushes to the split supersymmetry 
\cite{ArkaniHamed:2004fb,Giudice:2004tc,ArkaniHamed:2004yi} 
scenario, due to which the second 
problem is automatically solved: threshold corrections to the fermion 
masses are suppressed by powers of the large sfermion masses. Obviously, 
as usually in the split SUSY scenario, there are no $d=5$ proton decay operators, 
as well as no dangerous flavour changing neutral currents in the SM. 
The only solution is possible however for relatively small values of the GUT scale, 
very close to the experimentally allowed value. If this model turns out to be correct, 
the next generation of proton decay searches must be unavoidably successful
\footnote{This conclusion may change at the 2-loop order: after all, the fit determines 
only $\log{M_{GUT}}$, while the proton decay 
lifetime is proportional to $M_{GUT}^4$.}! 
Not only this: the presence of $d=6$ operators only and the known 
fermion flavour mixing matrices obtained from the fit make proton decay 
well known in all its channels. The model thus connects proton decay with 
neutrino masses, one of the ultimate goals of all grand unified theories.

The paper is written as follows: in section \ref{minso10} we will shortly review 
the minimal renormalizable supersymmetric  SO(10) model, as well as set up 
the notation to be used later. In section \ref{numass} we will first summarize the 
(well known) main problem in the fitting so far: the small neutrino overall scale. 
We will then propose a general strategy of how to remedy that problem in a phenomenologically 
acceptable way. 
First, from gauge coupling unification and $d=6$ proton decay constraints the 
allowed parameter space will be deduced. Second, we will scan that 
parameter space by performing a general 
fit of all fermion parameters except the neutrino overall scale. In this way we will 
see how close this scale can be to the one in agreement with experiments (in all interesting fits 
the neutrino spectrum will turn out to be normally hierarchical) for acceptably small $\chi^2$ of 
the general fit. We will conclude that a factor of about $5$ will be missing 
in order to get the correct neutrino mass scale in a scenario with low energy supersymmetry, 
thus confirming the conclusions of \cite{Aulakh:2005bd,Bajc:2005qe,Aulakh:2005mw,Bertolini:2006pe}. 
This will push us to consider in section \ref{splitSUSY} a scenario with a different 
sparticle spectrum: instead of low energy supersymmetry we will allow the 
sfermion masses to differ from the gaugino and higgsino ones. 
This time our strategy will yield a successful numerical fit, predicting observable 
proton decay rates, the supersymmetry breaking 
scale to be around $10^{13}$--$10^{14}$\,GeV with gauginos and higgsinos in the 
100\,TeV region and leptonic mixing angle $\sin \theta_{13} \approx 0.1$. 
A theoretically interesting outcome of this analysis is that neutrino masses 
get the main contribution from the type II seesaw mechanism. 
Finally we will outline the main results in section \ref{conclusions} and describe 
some further work to be done.

\section{The minimal SO(10)}
\label{minso10}

The minimal supersymmetric SO(10) contains on top of the usual three 
generations of spinorial $16_F$ also the 
$210_H$, $126_H$, $\overline{126}_H$ and $10_H$ Higgs representations. 
The renormalizable superpotential $W$ is
\begin{eqnarray}
\nonumber
W&=&16_F ( Y_{10} 10_H +  Y_{\overline{126}} \overline{126}_H) 16_F  \\ \nonumber
&&+\frac{m}{4!}  210_H^2 + \frac{\lambda}{4!} 210_H^3
+ \frac{M}{5!}  126_H \overline{126}_H + \frac{\eta}{5!}  126_H 210_H \overline{126}_H \\ 
&&+ 
m_H {10_H}^2 +\frac{1}{4!} 210_H 10_H 
(\alpha {126_H}+\overline{\alpha} \overline{126}_H)\, ,
\end{eqnarray}
where $Y_{10}$ and $Y_{\overline{126}}$ are the two complex symmetric Yukawa 
matrices of the theory. 

The above Lagrangian will be used for the determination of the mass 
spectrum needed for the renormalization group equation running of gauge and Yukawa coupling constants on one side and for the derivation 
of the relevant mass matrices needed for the fitting of the light fermion 
masses and mixing parameters at the GUT scale on the other side.

For the first one of these two topics, i.e., the mass spectrum, we refer the reader 
to the existent literature~\cite{Fukuyama:2004xs,Bajc:2004xe,Aulakh:2004hm} 
(for very recent reanalysis see \cite{Malinsky:2008gq,Aulakh:2008xx}). 
The upshot of those studies is that after a required fine-tuning the Higgs sector 
has only eight real parameters:
\begin{equation}
m,\ \alpha,\ \overline{\alpha},\ |\lambda|,\ |\eta|,\
\phi=\mbox{arg}(\lambda)=-\mbox{arg}(\eta),\  
x=\mbox{Re}(x)+ i \mbox{Im}(x).
\end{equation}
Here we closely follow the notation advocated in \cite{Bajc:2005qe}.
Once these parameters are given the mass spectrum of the Higgs fields is completely determined and can be found numerically. (We set $\phi=0$ for simplicity in what follows.) These parameters also specify all the vacuum expectation values (VEVs) of the theory and hence, besides Yukawa couplings $Y_{10}$ and $Y_{\overline{126}}$, directly affect the predictions for the fermion masses and their mixing parameters. 

Let us now set up the notation regarding the fermion mass matrices 
following \cite{Bajc:2005qe}. All we will need are two matrix sum-rules:
\begin{eqnarray}
\label{charged}
M_u&=&\frac{N_u}{N_d} \tan\beta [ (1+\xi(x))M_d - \xi(x) M_e ] ,\\
\nonumber\\
\label{neutral} 
M_n&=&
\frac{v}{ m} \frac{\sin^2\beta}{\cos\beta } 
\alpha \sqrt{\frac{|\lambda|}{|\eta|}} 
\frac{N_u^2}{N_d} \left[ m_{I}  \, f_I (x)+m_{I\!I} \, f_{I\!I} (x)
\right],
\end{eqnarray}
where $N_u$ and $N_d$ as well as all the ratios of polinomials of $x$, i.e., $f_I$, $f_{I\!I}$ and $\xi$, are specified in Ref.~\cite{Bajc:2005qe}. $\tan\beta$, another parameter of the model, is the ratio of up-type and down-type VEVs of the MSSM-like Higgs fields and $v$($=174$\,GeV at the $M_Z$ scale) is the scale of SU(2) breaking. $m_{I}= M_e(M_d-M_e)^{-1} M_e-6\xi M_e+9 \xi^2 (M_d-M_e)$ ($m_{I\!I}= M_d-M_e$) is proportional to the type I (type II) seesaw contribution to the light neutrino mass matrix. We remind the reader that type I seesaw is mediated 
by the right-handed neutrinos \cite{Minkowski:1977sc,Yanagida:1979as,GellMann:1980vs,Glashow:1979nm,Mohapatra:1979ia}, while 
the type II seesaw comes from the triplet VEV \cite{Magg:1980ut,Schechter:1980gr,Lazarides:1980nt,Mohapatra:1980yp}. $M_{u\,(d)}$ and $M_{e\,(n)}$ are $3 \times 3$ complex symmetric mass matrices of up (down) quarks and charged (neutral) leptons, respectively. Eqs.~(\ref{charged}) and~(\ref{neutral}) are crucial to establish whether a viable description of known fermion masses, mixing angles and phases at the GUT scale is possible. We will accordingly use them whenever we perform numerical fits. 

\section{The neutrino mass scale}
\label{numass}
The crucial question is whether this model can generate heavy enough light neutrino masses without being in conflict with other phenomenological constraints.  

\subsection{Maximization of the neutrino mass scale}

This issue is especially easy to illustrate in the context of a dominant type II seesaw mechanism. We accordingly focus on the magnitude of the  factor that multiplies the mismatch between the down-quark and charged lepton mass matrices, i.e., $M_d-M_e$, and thus controls the strength of the type II contribution 
\begin{equation}
\label{FII}
F_{I\!I}=\frac{v}{ m} \frac{\sin^2\beta}{\cos\beta } 
\alpha \sqrt{\frac{|\lambda|}{|\eta|}} 
\frac{N_u^2}{N_d} |f_{I\!I}(x)|.
\end{equation}
We want it to be as large as possible and still phenomenologically viable.
We will comment on the type I contribution in concrete examples to show that due to gauge coupling unification considerations it does not play a decisive role in establishing the correct neutrino mass scale.

We first note that $F_{I\!I}$ has to be at least of the order of $0.2 \times 10^{-9}$ for the model to successfully reach a lower bound on the heaviest light neutrino mass that is approximately $0.05 \times 10^{-9}$\,GeV~\cite{Strumia:2006db}. This is due to the well-known fact that $b$-$\tau$ unification happens rather naturally in supersymmetric theories at the GUT scale. We find this to still be true even after we properly incorporate the intermediate scales in the running of fermion masses in our numerical studies when the SUSY scale is low. Namely, the $b$ and $\tau$ massess that are of the order of $1$\,GeV defer from each other by not more than 25\,\% for the allowed values of $\tan \beta$ and hence lead to a cancellation in the $33$ element of the $M_d-M_e$ matrix. (This sort of cancellation is required in order to get a large atmospheric neutrino mixing angle~\cite{Bajc:2001fe,Bajc:2002iw,Bajc:2004fj}.) Therefore $F_{I\!I}$ has to be about a factor of $4$ bigger than one would naively expect in order to get a good fit of the overall neutrino scale. 

Clearly, a large enough $F_{I\!I}$ prefers a low enough $m$ barring some special cases when 
$|f_{I\!I}(x)|$ blows up. On the other hand, $m$ is proportional to the GUT scale which is bounded from below by the experimental limits on proton decay.  Therein lies the crux of the neutrino scale problem. Due to proton lifetime limits there exists a lower bound on $m$ that implies an upper bound on the attainable neutrino scale which, in general, tends to be too low to accommodate the experimental data on neutrino oscillations. 

To address this problem we use the current experimental bounds on the partial proton decay lifetimes to establish the phenomenologically viable lower limit on $m$. This allows us to establish an upper bound on $F_{I\!I}$---and hence an upper bound on the neutrino scale---that the theory can generate for particular values of $\alpha$, $\overline{\alpha}$, $|\lambda|$, $|\eta|$ and $x$ through the type II contribution while being certain that proton is stable enough. We find that the following conservative limit on $m$ applies
\begin{equation}
\label{m}
m> 5 \times 10^{15}\,{\rm GeV} |\lambda| \sqrt{\frac{A_S}{\pi}} \left(4 \left|\frac{2 x^2+x-1}{x-1}\right|^2 + 2 \left|\frac{2 x (2
   x^2+x-1)}{(x-1)^2}\right|^2 \right)^{-1/2}.
\end{equation}
For details of this derivation and notation we refer the reader 
to Appendix~\ref{ProtonDecay}. $A_S$ is taken to be a common value of the short distance enhancement factors of the relevant $d=6$ proton decay operators between the GUT scale and $M_Z$. These are calculable for a given mass spectrum of the theory. We take $A_S$ to be $2.5$ which overestimates the values obtained from running by 10--25\,\%.
Whenever we use $F_{I\!I}$ as defined in Eq.~(\ref{FII}) with $m$ explicitly replaced with the lower bound from Eq.~(\ref{m}) we refer to it as $F^{\rm{max}}_{I\!I}$.

The last step allows us to systematically maximize $F_{I\!I}$ in order to find a region where it exceeds the desired value for a successful description of the neutrino data. Since $F^{\rm{max}}_{I\!I} \sim 1/\sqrt{\lambda}$, a small value of $\lambda$ is preferred. If and when such a region is found we can further check whether the gauge coupling unification takes place there. If it does, we generate the appropriate mass matrices by propagating the experimentally determined fermion masses from $M_Z$ to the GUT scale taking care of all the intermediate states, and use these values to see if a successful fit of fermion masses and mixing parameters is indeed possible. This procedure guarantees a self-consistent check of the viability of the theory. (A similar 
line of approach has been taken in an SO(10) model with radiatively induced fermion mass hierarchy~\cite{Barr:2007ma} very recently~\cite{Barr:2008kg}.)

\begin{figure}[th]
\begin{center}
\includegraphics[width=5in]{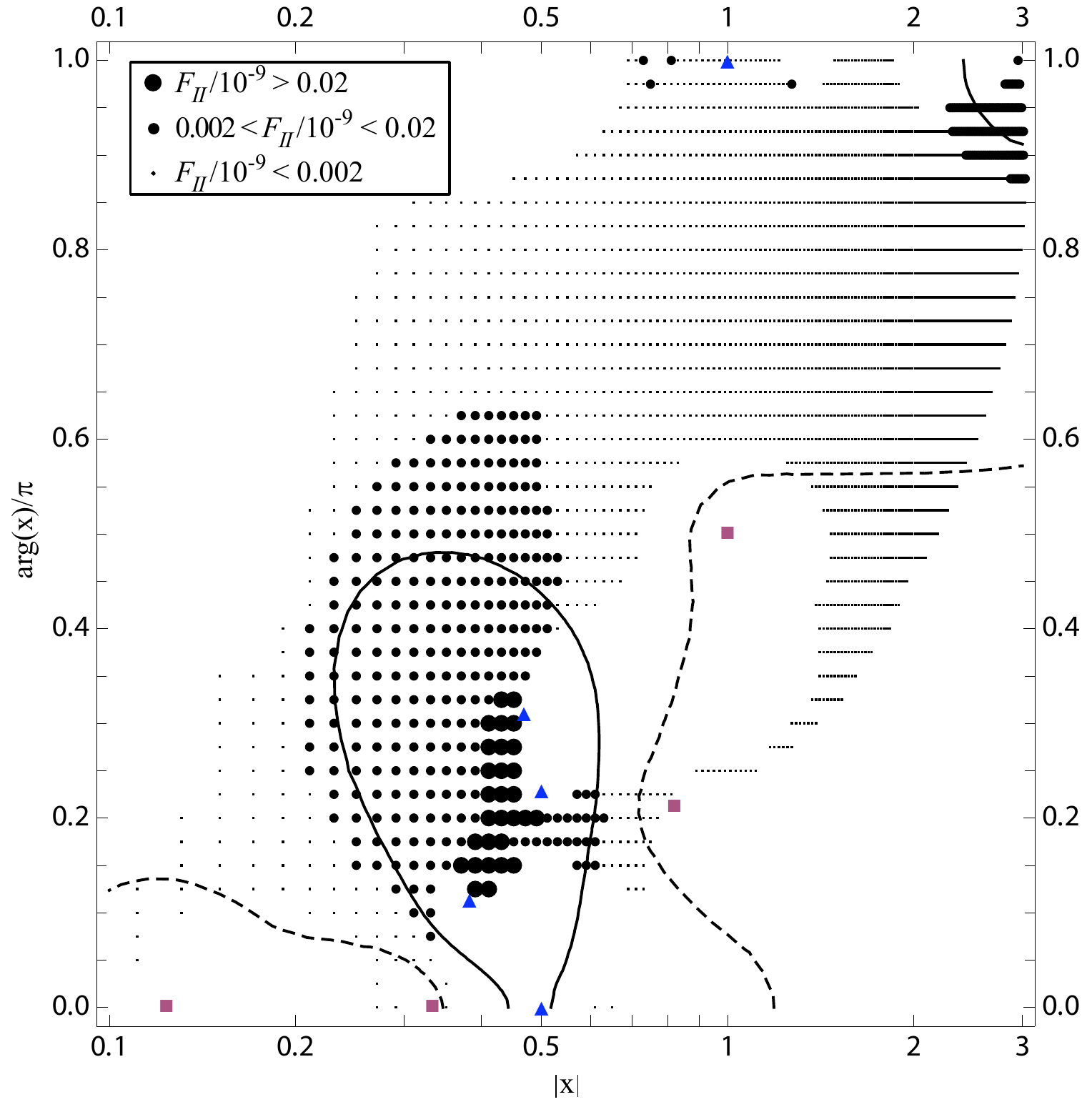}
\end{center}
\caption{\label{figure:FII} Lines of constant $F^{\rm{max}}_{I\!I}/10^{-9}$ for $\eta=\lambda=\alpha=\overline{\alpha}=1$ and $\tan \beta = 50$ and the corresponding viable gauge coupling unification points. Solid (dashed) line corresponds to $F^{\rm{max}}_{I\!I}/10^{-9}$ of $0.02$ ($0.002$). Dots of varying sizes mark $x$ values for which successful one-loop gauge coupling unification takes place for a fixed SUSY scale of $1$\,TeV. Their size indicates the magnitude of $F_{I\!I}/10^{-9}$. Large squares (triangles) mark the singular points of $f_{I}$ ($f_{I\!I}$).}
\end{figure}

Before we optimize $F_{I\!I}$, we first plot the lines of constant $F^{\rm{max}}_{I\!I}$ for one particular set of values of the relevant parameters ($\eta=\lambda=\alpha=\overline{\alpha}=1$, $\tan \beta = 50$) in Fig.~\ref{figure:FII} to show how close to the right neutrino scale one usually gets. Note that we always run $\tan \beta$ ($v$) from the SUSY ($M_Z$) scale to the GUT scale using the relevant 
equations to perform the numerical fit as well as to evaluate $F_{I\!I}$. However, whenever we specify $\tan \beta$ throughout this paper, we specify its value at the relevant SUSY scale unless stated otherwise.

Lines of constant $F^{\rm{max}}_{I\!I}/10^{-9}$ shown in Fig.~\ref{figure:FII} correspond to $0.02$ (solid line) and $0.002$ (dashed line). Recall that these contours reflect the actual upper bounds on $F_{I\!I}$ that are in accord with proton decay limits. Clearly, only in a very narrow region $F^{\rm{max}}_{I\!I}$ is within a factor of ten away from the preferred value. We can further ask if indeed unification takes place in that particular region and check whether it confirms our estimate for the mismatch between the required value for $F_{I\!I}$ and the actual value for $F_{I\!I}$. The dots  in Fig.~\ref{figure:FII} mark places where successful one-loop gauge coupling unification takes place assuming that the GUT scale $M_{GUT}$ is given by the scale of the lightest proton decay mediating $(X,Y)$ gauge bosons $M_{(X,Y)}$. We plot only the points that are in agreement with the experimental limits on proton decay lifetimes and yield a perturbative unification below the Planck scale. There are three different dot sizes; the largest dots represent unification scenario where $F_{I\!I}/10^{-9} > 0.02$, medium size dots correspond to $0.002 < F_{I\!I}/10^{-9} < 0.02$ and the smallest dots are for $F_{I\!I}/10^{-9} < 0.002$. Here we note that not even one point corresponds to the case where $F_{I\!I}/10^{-9}$ reaches or exceeds $0.2$. In other words, we can be rather certain that no successful fermion mass fit with dominant type II scenario for neutrino masses can be found for these particular values of model parameters due to a too small neutrino mass scale. 

As far as the pure type I contributions are concerned we show in Fig.~\ref{figure:FII} that all the singular points of $f_{I}$, represented with squares, do not overlap with the viable unification points. In other words, whenever type I contribution is potentially significant for light neutrino masses, unification does not happen.

\subsection{Constraints from unification and proton decay}

The exact procedure that we use to check for unification of gauge couplings for data in Fig.~\ref{figure:FII} is as follows. First, we fix the SUSY scale $M_{SUSY}=1$\,TeV. We further set 
$\eta=\lambda=\alpha=\overline{\alpha}=1$ and then vary $x$ and $g_{GUT}$, where $g_{GUT}$ is the gauge coupling at the GUT scale. Once all these parameters are given we numerically determine the masses of the superheavy fields, including the gauge ones, in arbitrary units of $m$. Then we define the following coefficients
\begin{equation}
B_i = \sum_{I} b_{i}^I r_{I}, \qquad r_I=\frac{\ln
M_{GUT}/M_{I}}{\ln M_{GUT}/M_{SUSY}}, \qquad (0 \leq r_I \leq 1),
\end{equation} 
where $b^I_i$, $i=1,2,3$, are the usual one-loop coefficients of $\beta$ functions of the $I$th threshold of the $i$th gauge coupling. Note that at this stage the only unknown is $\ln M_{GUT}/M_{SUSY}$. Finally, we solve for it using the difference between the one-loop equations for the running of $\alpha_1$ and $\alpha_2$ from $M_{SUSY}$ to $M_{GUT}$ and check that this value indeed generates unification and that the inferred gauge coupling at the GUT scale matches the input value for $g_{GUT}$. This we do by requesting that $(B_2-B_3)/(B_1-B_2)$ and $\alpha_{GUT}=g_{GUT}^2/(4 \pi)$ defer by less than 4\,\% from the central values of $(\alpha^{-1}_2-\alpha^{-1}_3)/(\alpha^{-1}_1-\alpha^{-1}_2)$ and $(\alpha_3^{-1}-B_3 \ln M_{GUT}/M_{SUSY}/(2 \pi) )^{-1}$, respectively. We introduce the $4$\,\% factors to reflect the fact that these are one-loop considerations for a single SUSY scale only and for central values of gauge couplings. We take $\alpha_3 = 0.0895$, $\alpha_2 = 0.0326$ and $\alpha_1 = 0.0174$ to be the input values at the SUSY scale of $1$\,TeV. These reflect the two-loop running effects from $M_Z$ to $1$\,TeV. We finally check that $\alpha_{GUT}$ is perturbative, proton decay constraints are satisfied and GUT scale itself is below the Planck scale. To check the proton decay viability we look at $p \rightarrow \pi^0 e^+$ channel and the $d=6$ operator contribution. To accurately evaluate the prediction for $p \rightarrow \pi^0 e^+$ we numerically find the relevant short distance coefficients, $A_{S\,L}$ and $A_{S\,R}$, of the $d=6$ operators (see Appendix~\ref{ProtonDecay}) taking care of all intermediate scales using the results of Refs.~\cite{Buras:1977yy,Wilczek:1979hc,Ellis:1979hy,Ibanez:1984ni,Munoz:1986kq} and use $g_{GUT}$, $M_{(X,Y)}$ and $M_{(X',Y')}$ as deduced from unification.

In Figs.~\ref{figure:tan30} and \ref{figure:tan50} we show the results of the fermion mass fit at the GUT scale when we actually run the charged fermion masses, the Cabibbo-Kobayashi-Maskawa (CKM) parameters and $v$ from $M_Z$ to $M_{GUT}$ and $\tan \beta $ from $M_{SUSY}$ to $M_{GUT}$ for $\tan \beta = 30$ and $\tan \beta = 50$, respectively. When we run the charged fermion masses, the CKM parameters, $v$ and $\tan \beta$ we use the knowledge of the full mass spectrum of the theory to properly include all intermediate scales at the one-loop level. We do not run the light neutrino masses and the known angles of the Pontecorvo-Maki-Nakagawa-Sakata (PMNS) mixing matrix. Again, the dots represent the points where gauge coupling unification takes place when $\eta=\lambda=\alpha=\overline{\alpha}=1$. The level of shading specifies the goodness of the fit, i.e., the range of values of $\chi^2$, which exhibits clear dependence on $\tan \beta$. Since we already know that the correct mass scale for the neutrinos cannot be reached, we fit beside the charged fermion masses, CKM parameters, solar and atmospheric angles of the PMNS matrix, one particular ratio $m_2/m_3$ of neutrino masses and then evaluate the mismatch between the mass of the heaviest light neutrino $m_3$ as inferred from the fit and lower bound on neutrino mass scale that we take to be $0.05$\,eV. For the details of the fitting procedure we refer the reader to Appendix~\ref{FittingProcedure}. We find $(0.05\,{\rm eV}/m_3) \geq 57$ for numerical fits for which $\chi^2/\mathrm{d.o.f.}<10/13$. This not only confirms our initial estimate that the theory fails to accommodate the relevant neutrino scale but gives us an accurate mismatch factor for the $\eta=\lambda=\alpha=\overline{\alpha}=1$ case with low SUSY scale.
 
\subsection{The failure of the low energy supersymmetric case}
 
The main lessons to be taken from this example are the following. Firstly, $F_{I\!I}$ when combined with the limit on $m$ as given in Eq.~(\ref{m}) provides direct means to accurately estimate whether a successful description of the light neutrino mass scale can be achieved. This should be the starting point of any particular scan of the parameter space of the theory. 

\begin{figure}[th]
\begin{center}
\includegraphics[width=5in]{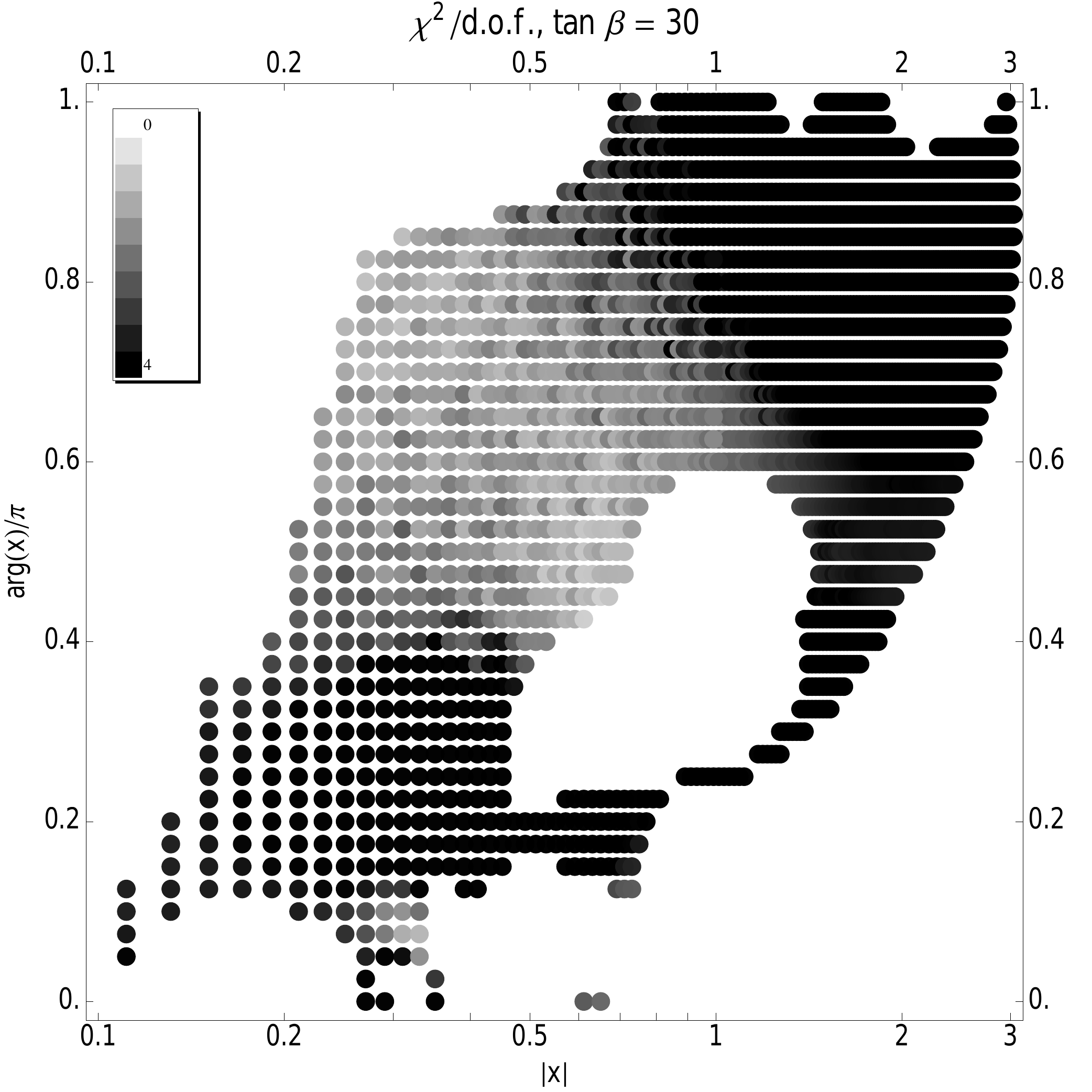}
\end{center}
\caption{\label{figure:tan30} Result of the numerical fit of charged fermion masses, $m_2/m_3$, CKM parameters, and solar and atmospheric angles for $\tan \beta = 30$. Dots correspond to successful gauge coupling unification
at the one-loop level for central values of low-energy
observables when $\eta=\lambda=\alpha=\overline{\alpha}=1$ and $M_{SUSY}=1$\,TeV. Their shading describes the goodness of the fit expressed through the magnitude of the $\chi^2$ function. Notice that we do not perform the fit in the white region without dots, because of the absence of unification and/or too fast proton decay.}
\end{figure}

\begin{figure}[th]
\begin{center}
\includegraphics[width=5in]{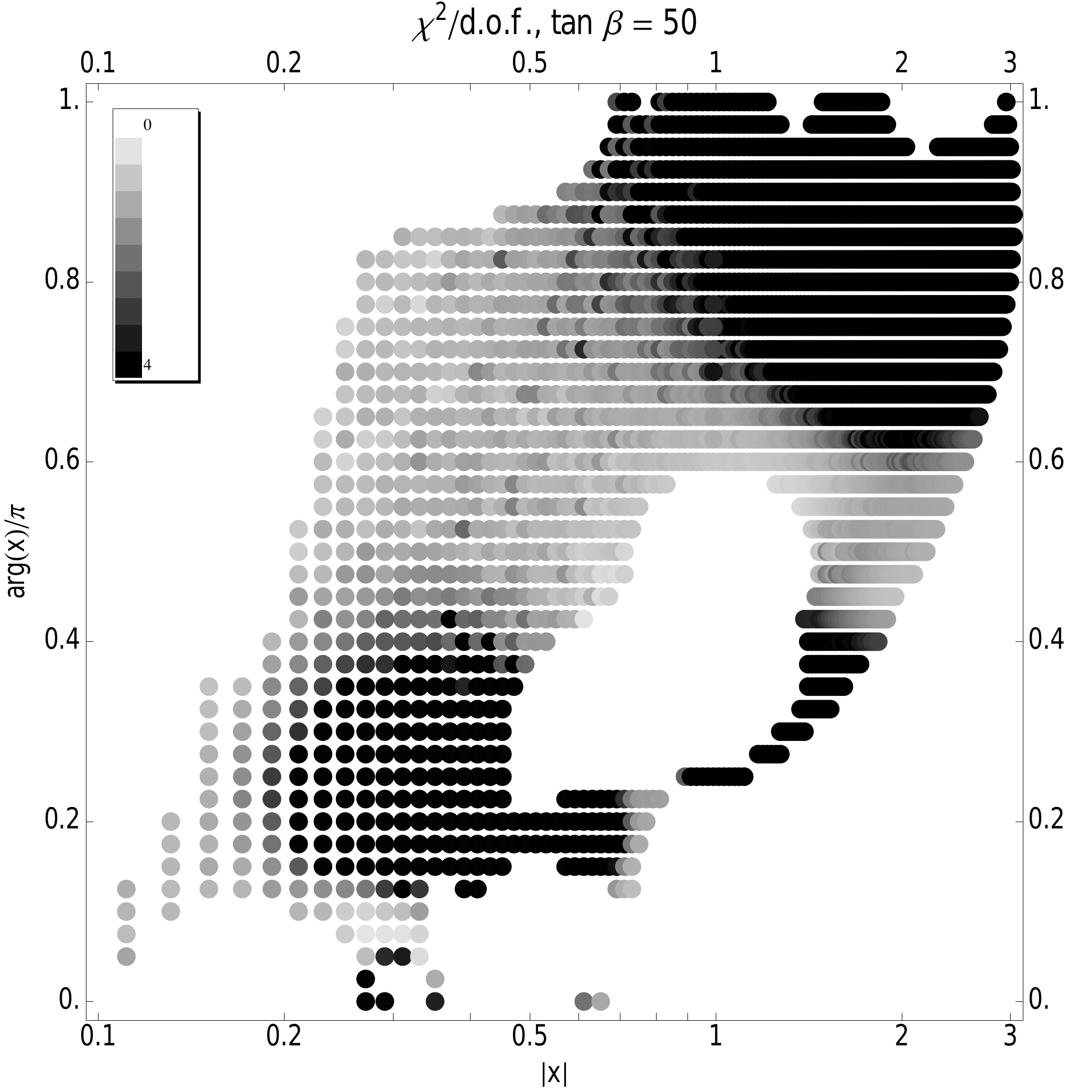}
\end{center}
\caption{\label{figure:tan50} Goodness of the fit of fermion masses and mixing angles for $\tan \beta = 50$, when  $\eta=\lambda=\alpha=\overline{\alpha}=1$ and $M_{SUSY}=1$\,TeV.}
\end{figure}

Secondly, $m$ is just another parameter in the theory and as such must be allowed to vary as long as inequality in Eq.~(\ref{m}) is satisfied in order to cover all possible unification scenarios. It actually varies from $4 \times 10^{14}$\,GeV to $4 \times 10^{17}$\,GeV for unification points shown in Figs.~\ref{figure:FII},~\ref{figure:tan30} and \ref{figure:tan50}. Fixing $m$ to a particular value would reproduce only a portion of the available parameter space. 

Thirdly, charged fermion masses vary drastically from point to point. The reason for this is 
that new (compared to the desert scenario) heavy states change the relevant renormalization group equations (RGEs). When we propagate the Yukawa couplings we neglect for simplicity the direct influence of GUT states threshold corrections, but take them consistently into account in the RGEs 
for the gauge couplings. This means that Yukawa couplings at the GUT scale differ from the ones 
obtained in the desert like MSSM scenario only due to a change in the values of the gauge couplings.
We will comment on this later on. 
In Table~\ref{table:range} we present the ranges of the central values of masses of quarks and charged leptons after we extrapolate them to the GUT scale for the unification points presented in Fig.~\ref{figure:FII} and confront them with the central values when we assume MSSM-like scenario with $\tan \beta = 30$, $M_{GUT} = 2 \times 10^{16}$\,GeV and $M_{SUSY} = 1$\,TeV. 
This, in turn, affects the numerical fit which depends on $\tan \beta$ for two reasons: 
it affects the RGEs as explained above and directly enters mass sum rules 
(\ref{charged})-(\ref{neutral}). 

Finally, it is clear that the unification consideration is crucial since unification happens in a rather narrow region of $x$. For example, all regions that are potentially viable in describing neutrino scale through the type I seesaw do not overlap with the regions where couplings unify when $\eta=\lambda=\alpha=\overline{\alpha}=1$ and are thus ruled out prior to any numerical fitting.

\begin{table}[htdp]
\begin{center}
\begin{tabular}{ c || c | c | c | c | c | c | c | c | c |}
   & $m_u$ & $m_c$ & $m_t$ & $m_d$ & $m_s$ & $m_b$ & $m_e$ & $m_\mu$ & $m_\tau$ \\ \hline
  min  &  $.00053$ & $.210$ & $86.7$ & $.0012$ & $.0213$ & $1.17$ & $.000341$ & $.0720$ & $1.28$\\
  max  & $.00060$ & $.237$ & $93.2$ & $.0014$ & $.0240$ & $1.29$ & $.000367$ & $.0775$ & $1.37$\\
  \hline 
  MSSM  &  $.00050$ & $.198$ & $75.7$ & $.0011$ & $.0202$ & $1.07$ & $.000357$ & $.0754$ & $1.34$
\end{tabular}
\end{center}
\caption{Range of central values of fermion masses in GeV units at the GUT scale when $\eta=\lambda=\alpha=\overline{\alpha}=1$ and $\tan \beta=30$ for different values of $x$ where unification takes place. We also present the two-loop level running masses at the GUT scale in the MSSM-like setup when $M_{SUSY}=1$\,TeV and $M_{GUT}=2 \times 10^{16}$\,GeV.}
\label{table:range}
\end{table}

Clearly, the theory fails to describe the current data on fermion masses and their mixing angles with generic parameters. We thus resort to a different strategy that we already alluded to at the beginning of this section. Namely, we first maximize $F^{\rm{max}}_{I\!I}$ in terms of $x$, $\eta$, $\lambda$, $\alpha$ and $\overline{\alpha}$ in order to make sure that $F^{\rm{max}}_{I\!I}$ is large enough for reasonable value of $\tan \beta$ to yield the correct neutrino mass scale. For that purpose we find suitable values of $\eta$, $\lambda$, $\alpha$ and $\overline{\alpha}$ by varying them in the following range: $0.03$--$7$. Note that this fixes the value of $m$ and thus sets the overall scale of the theory that is relevant for proton decay in terms of the scale that is relevant for the description of light neutrinos. Then we check whether unification of gauge couplings takes place. This we do by varying $g_{GUT}$ within a given perturbative range to try to simultaneously satisfy all three RGEs of the gauge coupling constants while assuming that $M_{SUSY}=1$\,TeV. Again, once we know $x$, $\eta$, $\lambda$, $\alpha$, $\overline{\alpha}$, $m$ and $g_{GUT}$ we have a full knowledge of the mass spectrum of the theory except $M_{SUSY}$. Finally, if and when unification works, we run the charged fermion masses, CKM parameters, $\tan \beta$ and $v$ to the GUT scale and perform the numerical fit as described in detail in Appendix~\ref{FittingProcedure} to check the viability of the theory. 

Although the above analysis is an improvement of the existing studies \cite{Aulakh:2005bd,Bajc:2005qe,Aulakh:2005mw,Bertolini:2006pe}, we still do not find a satisfactory solution with a single, low SUSY scale. In fact, $m_3$ comes out short by a factor of 5 or more with respect to the minimal experimental value, primarily due to the fact that unification with a low SUSY scale does not allow for sufficiently small values of $\lambda$.

\section{Split SUSY case}
\label{splitSUSY}
 
To reconcile the sufficiently small values of $\lambda$ with unification we split $M_{SUSY}$ into two scales. The first scale stands for the common scale $M_{1/2}$ of gauginos and higgsinos while the second one describes the common scale $M_{\tilde f}$ of matter field superpartners. We now simply solve the three relevant RGEs for $M_{1/2}$, $M_{\tilde f}$ and $g_{GUT}$ to insure exact gauge coupling unification at the one-loop level.  Within this scenario we find a multitude of solutions for various values of the parameters for which $F^{\rm{max}}_{I\!I}$ looks promising enough to attempt a numerical fitting of fermion masses and mixing parameters. As it turns out, these solutions are always of the split SUSY type with light gauginos and higgsinos in the 100\,TeV range and superheavy squarks and sleptons that are close to the GUT scale (but still low enough to allow a supersymmetric treatment of the potential and of the mass spectrum). We accordingly perform the running of the Yukawa couplings from $M_Z$ to $M_{GUT}$ taking care of the split supersymmetry effects as prescribed in Ref.~\cite{Giudice:2004tc} in the presence of intermediate scales for various values of $\tan \beta$ and execute a numerical fit of charged fermion masses, CKM parameters, two PMNS mixing angles and one particular ratio of light neutrino masses $m_2/m_3$.

We find a solution that yields both successful unification and satisfactory $\chi^2$ which we explicitly spell out in Table~\ref{tab:n37}. There we also specify our input values for charged fermions and CKM parameters at $M_Z$. These $M_Z$ values are updates of results already presented in Ref.~\cite{Dorsner:2006hw} to reflect the results of Ref.~\cite{PDG}. The numerical fit determines also leptonic angle $\sin \theta_{13} (\equiv s^{PMNS}_{13})$, the only angle in the PMNS matrix that is yet to be determined.  

\begin{table}[th]
\begin{tabular}{|c|r|rrr|}
\hline
observable \qquad& data at $M_Z$ & FIT & RGE & pull \\
\hline
$m_e$(MeV) & $0.4866613 $ & $0.42279$ & $ 0.42279$ & $  $ \\
$m_\mu$(GeV) & $0.10273 $ & $0.08925$ & $ 0.08925$ & $  $ \\
$m_\tau$(GeV) & $1.746 $ & $1.534$ & $ 1.534$ & $  $ \\
$m_u$(MeV) & $1.6 $ & $0.56 $ & $ 0.54 $ & $ +0.069  $ \\
$m_c$(GeV) & $0.628 $ & $0.218$ & $ 0.214$ & $ +0.182 $ \\
$m_t$(GeV) & $171.5 $ & $72.0 $ & $ 67.4 $ & $ +0.675$ \\
$m_d$(MeV) & $3.5 $ & $0. 24$ & $ 1.2$ & $ -2.310$ \\
$m_s$(MeV) & $62 $ & $27.2$ & $ 21.8$ & $ +0.696$ \\
$m_b$(GeV) & $2.89 $ & $0.917$ & $ 0.910$ & $ +0.079 $ \\
$s^{CKM}_{12}$ & $0.2272 $ & $0.2423$ & $ 0.2272$ & $ +0.951$ \\
$s^{CKM}_{23}$ & $0.0422 $ & $0.0478$ & $ 0.0474$ & $ +0.199$ \\
$s^{CKM}_{13}$ & $0.00399 $ & $0.00447$ & $ 0.00448$ & $ -0.015$ \\
$\delta^{CKM}$ & $0.995 $ & $1.149$ & $ 0.995$ & $ +0.598$ \\
$s^{PMNS}_{12}$ & $ $ & $0.42$ & $ 0.55 $ & $ -1.091$ \\
$s^{PMNS}_{23}$ & $ $ & $0.55$ & $ 0.69 $ & $ -0.764$ \\
$s^{PMNS}_{13}$ & $ $ & $0.103$ & $  $ & $  $ \\
$m_2/m_3$ & & $ $ $0.178$ & $ 0.180 $ & $ -0.104 $ \\
\hline
\end{tabular}
\caption{\label{tab:n37}
Input (RGE) and output (FIT) parameters of the numerical fit at the GUT scale with $\chi^2/\rm{d.o.f.} =9.6/13$, $m_3 =0.049$\,eV, $m_2 =0.0087$\,eV and $m_1 =0.0012$\,eV. Unification takes place for $|x|=0.109$, $arg(x)/\pi=0.52$, $\alpha=1.26788$, $\overline{\alpha}=7$, $\eta=6.54112$, 
$\lambda=0.03$, 
$M_{1/2} =1.5 \times 10^{5}$ GeV, 
$M_{\tilde f}=9.0 \times 10^{13}$ GeV, 
$M_{GUT}=5.8 \times 10^{15}$ GeV 
and $g_{GUT}=1.3$ ($m=6.5\times 10^{13}$ GeV).  
The input data at $M_Z$ are also given in the second column.}
\end{table}

Our approach to the numerical fitting, as described in Appendix~\ref{FittingProcedure}, is statistical in nature. In view of that we cannot guarantee that our minimization procedure always finds a true global minimum of  $\chi^2$. In other words, there could still be some room for improvement as far as the numerical part of this study is concerned. Moreover, the complexity of the problem prevents us from scanning the whole available parameter space since we must always fix $x$, $\eta$, $\lambda$, $\alpha$, $\overline{\alpha}$, $m$, $g_{GUT}$ and $\tan \beta$ prior to numerical running and fitting of fermion masses and mixing parameters. Our main finding, though, remains: it is possible for the minimal renormalizable SO(10) to yield correct neutrino mass scale while being in agreement with the gauge coupling unification paradigm as well as proton decay constraints.

The minimal renormalizable SO(10) establishes a direct connection between the neutrino scale and the GUT scale. In fact we have used this connection as a guiding tool towards discovery of the viable parameter space. Since our solution requires a GUT scale that is rather low, any significant improvement with respect to the current experimental limits on the $p \rightarrow \pi^0 e^+$ lifetime would put serious constraint on this model. Of course, as already mentioned in the introduction, before reaching this conclusion, a two-loop check would be in order, due to the large sensitivity of the proton decay rate on the value of $M_{X,Y}$ and $M_{X',Y'}$. Be that as it may, we summarize in Table~\ref{tab:Proton} the  predictions of this model for the most relevant proton decay channels for the fit shown in Table~\ref{tab:n37}. We derive them using the flavour structure dependence of the relevant $d=6$ operators spelled out in Ref.~\cite{FileviezPerez:2004hn}.
\begin{table}[th]
\begin{tabular}{|l|r|l|r|}
\hline
& \multicolumn{2}{|c|}{Partial mean life ($10^{33}$\,years)} &\\
\hline
$p$ decay modes & Fit: Table~\ref{tab:n37} & Lifetime bounds:~\cite{PDG08} & Fraction ($\Gamma_i /\Gamma$)\\
\hline
$p \rightarrow \pi^0 e^+$ & $ 1.6 $ & $>1.6$ & $44.3$ \% \\
$p \rightarrow \pi^0 \mu^+$ & $ 70 $ & $>.473$ & $1.0$ \%\\
$p \rightarrow K^0 e^+$ & $ 442 $ & $>.150$ & $0.2$ \%\\
$p \rightarrow K^0 \mu^+$ & $ 15 $ & $>.120$ & $4.7$ \%\\
$p \rightarrow \eta e^+$ & $ 238 $ & $>.313$ & $0.3$ \%\\
$p \rightarrow \eta \mu^+$ & $ 238 $ & $>.313$ & $0.5$ \%\\
$p \rightarrow \pi^0 \bar{\nu}$ & $ 1.5 $ & $>.025$ & $49.0$ \%\\
\hline
\end{tabular}
\caption{\label{tab:Proton}
Predictions for partial lifetimes for the most significant $p$ decay modes for the fit shown in Table~\ref{tab:n37} and current bounds~\cite{PDG08} expressed in $10^{33}$\,years units. For the matrix element we take $\alpha=0.009$\,GeV$^3$~\cite{Tsutsui:2004qc}. The predicted associated branching ratios are also shown.}
\end{table}

For completeness we explicitly 
provide the mixing matrices $W_e$ and $V_q$, $\tan \beta$, and $v$ at the GUT scale as well as $A_{S\,L}$ and $A_{S\,R}$ coefficients for the successful fit defined in Table~\ref{tab:n37}. $W_e$, $V_q$, $A_{S\,L}$ and $A_{S\,R}$ are defined in Appendix~\ref{ProtonDecay}.
\begin{eqnarray}
W_e=\pmatrix{
-0.9989 + 0.0260\,i & -0.0288 + 0.0227\,i & +0.0136 - 0.0023\,i \cr
-0.0132 - 0.0348\,i & +0.8338 + 0.5478\,i & -0.0559 -0.0156\,i \cr
-0.0096 - 0.0080\,i & -0.0323 - 0.0486\,i & -0.7749 - 0.6293\,i},
\end{eqnarray}
\begin{eqnarray}
V_q=\pmatrix{
+0.8936 + 0.3778\,i & -0.2298 - 0.0771\,i & -0.0011 + 0.0043\,i \cr
-0.2386 - 0.0412\,i & -0.9648 - 0.0909\,i & -0.0438 + 0.0191\,i \cr
+0.0106 + 0.0002\,i & +0.0440 + 0.0159\,i & -0.9828 + 0.1785\,i}
\end{eqnarray}
\begin{equation}
\tan \beta=34.6, \qquad v=147.9\,\rm{GeV}, \qquad A_{S\,L}=2.0, \qquad A_{S\,R}=2.2.
\end{equation}
\noindent

One comment is needed before we end this section. 
As we said before, the Yukawa couplings were run without taking into account the 
corrections due to new couplings appearing above the heavy thresholds. For small enough 
values of these new couplings, the effect is negligible due also to the small mass ratio between 
heavy thresholds and $M_{GUT}$. But if the couplings are large enough (as they are here for 
example $\eta$ and $\bar\alpha$), one may worry that 
they would completely change the $M_{GUT}$ values of the light fermion masses and mixings 
that we then want to fit. We estimated this effect and found out that it cannot dominate the running. 
There are more arguments in favor of our claim. First of all, the corrections to the 1-loop RGE due to 
new interactions above heavy thresholds is due to the wave-function renormalization of the light 
Higgs fields. This means that it is approximately flavor independent for short running and thus have 
little impact on the quantities we are fitting. Secondly, most of the lighter fields among the heavy 
ones are coming from $210_H$. But the dangerous couplings can come from the operators 
$210_H 126_H\overline{126}_H$ (large $\eta$ coupling) or $210_H 10_H \overline{126}_H$ 
(large $\bar\alpha$ coupling). So the projections of fields coming from $126_H$ or $\overline{126}_H$ 
are typically small in the light eigenvector directions, so the effective couplings are much smaller 
than the original SO(10) ones, and thus the effect becomes 
negligible. The same can be said for the $\overline{126}_H$ contribution ($\alpha/\eta$ suppressed) 
in the MSSM 
Higgs  $H_{u,d}$ directions, which again causes a small coupling of these two fields to the 
weak triplet living in $210_H$. Finally, the contribution of the operators between SM singlets and the Higgs bilinears are small either because the components with lighter singlets have  small couplings 
or components with larger couplings are heavy enough.

\section{Conclusions}
\label{conclusions}

Of course we have not shown that the above solution is unique. In fact 
it is still possible that low energy supersymmetry is allowed by this minimal 
renormalizable SO(10): the unsuccessful fit to fermion masses and mixing parameters 
can be corrected by loops due to soft SUSY breaking terms. These same terms 
must be such to also reduce enough the typically large $d=5$ proton decay operators. 
What we found in this paper is one very simple among such 
solutions: split supersymmetry with a well determined scale of $10^{13}$--$10^{14}$\,GeV.
Such a scale is large enough to sufficiently change the value of $g_{GUT}$ in order 
to allow for a small enough value of $\lambda$ which, in turn, yields the 
correct neutrino mass scale.

Most predictions of this model are common to split supersymmetric models: 
no sfermions at the TeV scale, the LSP a natural dark matter candidate, 
long gluino lifetime and the absence of flavour changing effects as well as 
$d=5$ proton  decay operators. What is further predicted in our model is the exact 
form and magnitude of the $d=6$ proton decay operators: we know both the 
proton lifetime, together with all the branching ratios. Such a GUT is thus 
an example of a predictive theory of proton decay, similar to non-supersymmetric theories, 
where the flavour structure alone (and not unknown supersymmetry breaking parameters) 
governs the proton decay rates \cite{DeRujula:1980qc}. 
On top of that we have a good description of all the fermion masses and mixings 
in the SM, connecting thus neutrinos with charged leptons. As a bonus of this 
fit we find out that the yet unknown parameter $s^{PMNS}_{13}$ turns out to be 
relatively large: $0.10$. This cannot 
be treated as a prediction of the model, though. In fact, it is possible that other solutions exist. 
The same can be said for some of the other output numbers we have quoted above: for 
example, it is probably possible to find solutions with lower mass of some of the gauginos 
or higgsinos, if one allows them to be split. 
We confirm however that the low energy supersymmetric model with negligible 
corrections due to soft terms is not viable. 

We are aware of three weak points in the above analysis. 

First, some of the parameters 
are quite large (the biggest one is $\bar\alpha=7$) so that perturbativity may be 
lost. Although $(\bar\alpha/4\pi)^2< 1$, the large number of fields involved 
may make matters worse. 
Notice however that a large coupling is by itself not necessarily dangerous: to 
prove that the non-perturbative regime has been reached one would need 
to calculate the relevant processes (decay widths, cross sections) at the one loop level 
and compare them with the tree order expressions. Obviously such a huge calculation 
is beyond the scope of this paper. 

Second, as we already mentioned, we have not fully consistently run the Yukawa 
couplings; above the heavy thresholds only the gauge couplings RGEs were properly 
modified. Although we have estimated that the Yukawa couplings 
of our solution with split supersymmetry after running from $M_Z$ to $M_{GUT}$ 
do not get sizable corrections from the new operators above the heavy thresholds, 
it would be desirable to include these effects from 
the beginning. This would be needed among others to definitely prove that the 
low SUSY scenario is ruled out in this context. 
It is reassuring however, that in spite of the large couplings involved 
the actual corrections turn out to be small. This is a signal that the theory may still be 
in a perturbative regime, and that the first weak point mentioned above may not be 
crucial after all.

Third, the split supersymmetry scale $M_{\tilde f}$ 
is pretty large and exceeds the usual cosmological bounds 
\cite{Antoniadis:2004dt,Arvanitaki:2005fa}, although it is partially alleviated 
by the large gluino scale $M_{1/2}$. Such a situation is similar to the case of 
radiative induced seesaw scale proposed in \cite{Bajc:2004hr}: there also the split 
SUSY scale $M_{1/2}$ was predicted to be large, close to the GUT scale, although the 
reason was different (to prevent right-handed neutrinos from becoming too light). 
Here as there the solution to this problem could be a heavy enough gluino or a 
late inflation, so that gluinos are not produced later. This and the previous issues deserve 
a separate study, which we postpone for the future.

\section*{Acknowledgements}

It is a pleasure to thank Goran Senjanovi\' c for reading the manuscript, comments, 
discussions and encouragements, Charan Aulakh and Alejandra Melfo for correspondence 
and clarifying discussions on normalizations, and Thomas Schwetz for sharing data for 
comparison. I.D.\ would like to thank Pavel Fileviez P\' erez for discussion.
This work has been supported by the Slovenian Research Agency 
(B.B. and M.N.) and by the Marie Curie International Incoming Fellowship within  the $6^{th}$ European Community Framework Program (I.D.). 

\appendix
\section{Proton decay}
\label{ProtonDecay}
Here we derive a constraint on the mass spectrum of our model due to
the current experimental bounds on the partial proton decay
lifetimes. The idea is to replace the only relevant mass scale $m$ in the model with the mass scale that can be inferred from experimental data. For the moment we assume that the $d=5$ proton decay operators are subdominant to the $d=6$ ones. If that is the case, the most stringent limit comes from the $p \rightarrow \pi^0 e^+$ mode for which the current limit reads $\tau(p \rightarrow \pi^0 e^+) > 1.6 \times 10^{33}$\,years~\cite{PDG}. The model, on the other hand, yields the following width for that mode:
\begin{eqnarray}
\label{gamma}
\Gamma &=&\frac{m_p}{16 \pi f^2_{\pi}} \ A_L^2 \
|\alpha|^2  (1 + D + F)^2 \left[A^2_{S\,R}
\left|k_1^2 (W_e^\dagger)^{11} + k_2^2 (V_{q}^\dagger)^{11} (V_{q} 
W_e^\dagger)^{11})\right|^2\right. \nonumber\\
&&\left.+A^2_{S\,L} k_1^4 \left|
(W_e)^{1 1}+(V_{q})^{11} (W_e V_{q}^\dagger)^{1 1} \right|^2 \right],
\end{eqnarray}
where $W_e^\dagger=U_d^{\dagger}U_e$, $V_{q}=U_u^{\dagger}U_d$ and
 $k_{1\,(2)}=\sqrt{2 \pi \alpha_{GUT}}/M_{(X,Y)\,((X',Y'))}$.  $A_{S\,L\,(R)}$ give a 
leading-log renormalization of the relevant operators
from the GUT scale to $M_Z$. The QCD running below $M_Z$ is captured by the
coefficient $A_L$. We note that the mass of proton decay mediating gauge boson $M_{(X,Y)}$ in the minimal SO(10) is given by the following expressions~\cite{Aulakh:2004hm}
\begin{equation}
M_{(X,Y)}=m \frac{g_{GUT}}{|\lambda|} \sqrt{4 \left|\frac{2 x^2+x-1}{x-1}\right|^2 + 2 \left|\frac{2 x (2
   x^2+x-1)}{(x-1)^2}\right|^2}.
\end{equation}

To progress we assume that $A_{S\,L} \approx A_{S\,R}=A_{S}$, $U_d^{\dagger} = U_e$ and $M_{(X,Y)}\approx
M_{(X',Y')}$. In that case we infer the following conservative lower limit on $m$: 
\begin{equation}
\label{lowerm}
m> 5 \times 10^{15}\,{\rm GeV} |\lambda| \sqrt{\frac{A_S}{\pi}} \left(4 \left|\frac{2 x^2+x-1}{x-1}\right|^2 + 2 \left|\frac{2 x (2
   x^2+x-1)}{(x-1)^2}\right|^2 \right)^{-1/2}.
\end{equation}
To generate this result we use $m_p=938.3$\,MeV, $D=0.81$, $F=0.44$,
$f_{\pi}=139$\,MeV, $A_L=1.25$,
$|(V_{q})^{1 1}|=0.97377$ and
$\alpha=0.009$\,GeV$^3$~\cite{Tsutsui:2004qc}. 

Although the bound in Eq.~(\ref{lowerm}) has been obtained after some assumptions, we 
have checked at the end of the fit that the general expression for $\Gamma$ in Eq.~(\ref{gamma}) 
is smaller than the experimental bound.

\section{Fitting procedure in MSGUT}
\label{FittingProcedure}

Our approach to the numerical fitting of the relevant fermion masses and mixing parameters is straightforward. To perform the fit we first construct the minimization function of the input parameters, which we then numerically minimize using a so-called simulated annealing procedure~\cite{Press:2005}. Since all the mass matrices depend on two complex symmetric Yukawa matrices, i.e., $Y_{10}$ and $Y_{\overline{126}}$, we have 15 real parameters at our disposal to fit all the fermion masses and mixing parameters. In practice, we trade $Y_{10}$ and $Y_{\overline{126}}$ for the down quark and charged lepton mass matrices to arrive at the following sum rule~\cite{Bajc:2005qe} :
\begin{equation}\label{eqSumRule}
  V_q^T \hat M_u V_q = \frac{N_u}{N_d} \tan \beta \left [ \left(1+\xi(x)\right)\hat M_d - \xi(x) W_e^T \hat M_e W_e \right].
\end{equation}
Here we work in the mass eigenstate basis of down quarks. We implement congruent transformation with the following convention:
$$ \hat M_x = U_x^T M_x U_x,\, x = u,d,e,n,$$
where $ \hat M_x$ are diagonal, real and positive. The input parameters we vary in the fit are the three down quark masses in $\hat M_d$ and three angles and six phases that parametrize $W_e = U_e^\dagger U_d$. The charged leptons are basically taken out of the fitting procedure due to their small experimental error (we fix them to the central value at the GUT scale). Therefore we are left with 12 parameters altogether, 6 of which are phases. Together with the model parameters that determine the spectrum 
($\lambda$, $\eta$, $\alpha$, $\bar\alpha$, $x$, $m$, $g_{GUT}$) they completely determine the r.h.s.\ of Eq.~(\ref{eqSumRule}) and determine the up quark masses and the CKM mixing matrix, which is multiplied by five additional phases $V_q \equiv U_u^\dagger U_d = P_1 V_{CKM} P_2$, where 
$P_1={\rm diag}(e^{i\phi_1}, e^{i\phi_2}, e^{i\phi_3})$ and $P_2={\rm diag}(e^{i\beta_1}, e^{i\beta_2},1)$. (Similar expression holds also for the PMNS mixing matrix $V_\ell  \equiv U_e^\dagger U_n= K_1 V_{PMNS} K_2 $.) To obtain the neutrino masses and $V_{\ell}$, we simply plug $M_d$ and $M_e$ in the neutrino mass formula given in Eq.~(\ref{neutral}).

We now specify the pull, $\chi_i$, for each output value $p_i$ in the following way:
$$\chi_i = \frac{p_i - \tilde p_i}{f_i \tilde p_i}$$
where $\tilde p_i$ are masses and angles which have been run to the GUT scale. $f_i$ are the percentages that specify the errors and are listed in Table~\ref{tabFitVal}.
\begin{table}[htdp]
\begin{center}
\begin{tabular}{ c || c | c | c | c | c | c | c | c | c | c |}
        i   & 1 & 2 & 3 & 4 & 5 & 6 & 7 & 8 & 9 & 10 \\ \hline
  $p_i$ & $m_u$ & $m_c$ & $m_t$ & $m_d$ & $m_s$ & $m_b$ & $s^{CKM}_{12}$ & $s^{CKM}_{23}$ & $s^{CKM}_{13}$ & $\delta^{CKM}$ \\ \hline
  $f_i$ (\%) & $30$ & $10$ & $10$ & $35$ & $35$ & $10$ & $7$ & $4$ & $1$ & $20$
\end{tabular}
\end{center}
\caption{Fitted parameters of the charged fermions and their relative errors.}
\label{tabFitVal}
\end{table}
The minimization function is then a sum of squared pulls and consists of the quark and the neutral lepton parts:
$$\chi^2 = \chi^2_q  + \chi^2_\ell, \; \; \chi^2_q = \sum_{i = 1}^{10} \left( \frac{p_i - \tilde p_i}{f_i \tilde p_i} \right)^2. $$
Again, we fix the charged lepton masses at their central values at the GUT scale. Therefore their contribution to the $\chi^2$ is zero. Neutrinos, on the other hand, enter the fit in the following way. We notice that the fitting procedure results in a hierarchical spectrum of the light neutrinos, so we choose to fit the ratio of the two heaviest neutrinos $m_2/m_3$, which does not receive large RGE corrections. We further relax the maximal neutrino mixing angles, since the RGE corrections typically reduce them at the GUT scale in the case of hierarchical spectrum in both type I and type II case \cite{Antusch:2005gp, Schmidt:2007nq,Mohapatra:2005pw} to get:
\begin{eqnarray}
\chi^2_\ell &=& \left(\frac{m_2/m_3 - 0.18}{0.02} \right)^2 + \left(\frac{s^{PMNS}_{12} - 0.5}{0.15 \times 0.5} \right)^2 \\
&+& \left(\frac{s^{PMNS}_{23} - 0.6}{0.1 \times 0.6} \right)^2 + \left(\frac{s^{PMNS}_{13} - 0.17}{0.2 \times 0.17} \right)^2 \Theta \left( s^{PMNS}_{13} - 0.17 \right).\nonumber
\end{eqnarray}

As for the input values for
all charged fermion masses at $M_Z$, they are taken from~\cite{Dorsner:2006hw} and updated where needed to reflect the  results of Ref.~\cite{PDG}\footnote{After we finished the numerical work, the 
new version of PDG \cite{PDG08} has been released. We did not repeat the whole calculation again 
since we do not expect any sizeable difference in the final results.}
 as shown in Table~\ref{tab:n37}. In addition, we use the CKM mixing angles and the CP phase as inferred from the fit for the Wolfenstein parameters as given in~\cite{PDG}. For the gauge couplings at $M_Z$ we take $\alpha_3(M_Z) =
0.1176 \pm 0.0020$, $\alpha_2(M_Z) = 0.033816 \pm 0.000027$ and
$\alpha_1(M_Z) = 0.016949 \pm 0.000005$. These are then run at the one-loop level to the correct GUT scale for a given $\tan \beta$ where we perform the fit.

\end{document}